\documentclass[prl,aps,twocolumn,superscriptaddress,amsmath,amssymb]{revtex4}
\usepackage{amsmath}
\usepackage[english]{babel}
\usepackage{graphicx}
\usepackage{bm}
\usepackage{color}

\begin{document}

\title{Reply  to ``Comment on Yerin et al., Phys. Rev. Lett. 121, 077002 (2018), and Mironov et al., Phys. Rev. Lett. 109, 237002 (2012)'' by A. F. Volkov, F. S. Bergeret, and K. B. Efetov [1]}

\author{S. V. Mironov}
\affiliation{Institute for Physics of Microstructures, Russian Academy
of Sciences, 603950 Nizhny Novgorod, GSP-105, Russia}
\author{D. Yu. Vodolazov}
\affiliation{Institute for Physics of Microstructures, Russian Academy
of Sciences, 603950 Nizhny Novgorod, GSP-105, Russia}
\author{Y. Yerin}
\affiliation{Physics Division, School of Science and Technology, Universit\`{a} di Camerino Via Madonna delle Carceri 9, I-62032 Camerino (MC), Italy}
\author{A. V. Samokhvalov}
\affiliation{Institute for Physics of Microstructures, Russian Academy
of Sciences, 603950 Nizhny Novgorod, GSP-105, Russia}
\author{A. S. Mel'nikov}
\affiliation{Institute for Physics of Microstructures, Russian Academy
of Sciences, 603950 Nizhny Novgorod, GSP-105, Russia}
 \affiliation{Lobachevsky State University of Nizhny Novgorod, 23 Prospekt Gagarina, 603950, Nizhny Novgorod, Russia}
\author{A. Buzdin}
\affiliation{University Bordeaux, LOMA UMR-CNRS 5798, F-33405 Talence Cedex, France}
\affiliation{Department of Materials Science and Metallurgy, University of Cambridge, CB3 0FS, Cambridge, United Kingdom}
\affiliation{Sechenov First Moscow State Medical University, Moscow, 119991, Russia}

\maketitle

The first part of the comment \cite{Comm} concerns the possibility of the Fulde-Ferrell-Larkin-Ovchinnikov (FFLO) state realization in superconductor (S) / ferromagnet (F) bilayer pointed out in the works \cite{BVE_PRB, BVE_PRL} by the authors of the comment. In the case of the small thicknesses of S and F layers and effective averaging of magnetism and superconductivity, this possibility seems to be evident for us. In our paper, we do not pretend to introduce this idea, which by the way was mentioned in the works \cite{Proshin, Baladie} published before that of the authors of the comment \cite{BVE_PRB, BVE_PRL}. Therefore, the first part of the comment \cite{Comm} seems to be irrelevant to our works \cite{Mironov_2012, Mironov_2018}.

Further on the authors of the comment refute the criticism of their work \cite{BVE_PRB} citing from our paper \cite{Mironov_2012} the following sentence: {\it ``Recently, this observation has been questioned in several theoretical works [3–5] predicting the sign change in the London relation and an unusual paramagnetic response of the hybrid superconductor or ferromagnet (S/F) andsuperconductor or normal metal (S/N) systems''.} However, in this sentence there is absolutely no criticism, it is a simple recognition of the previous works and then the comment is also irrelevant to our paper \cite{Mironov_2012}.

Finally they cite from our paper \cite{Mironov_2018}: {\it ``It is exactly this FFLO instability which makes impossible to observe the global paramagnetism predicted in [34-36]. The latter paramagnetic state just does not correspond to the free energy minimum [37]''.} They note that in their work \cite{BVE_PRB} they are dealing with a local change of sign of the coefficient between the current density and vector potential. However, in \cite{BVE_PRB} on pages 8 and 9 and Figs. 9-11 they provide the total (integrated over the F film thickness) current which changes its sign. From a formal point of view we could, thus, consider the statement of the authors of \cite{Comm} that they do not calculate the integrated response in \cite{BVE_PRB} to be a sort of confusion. On the other hand, our citation of the work \cite{BVE_PRB} also is not completely correct. The important point is that in the above phrase from \cite{Mironov_2018} we have mixed up the total current in the F layer (which is induced by the internal magnetic induction in the ferromagnet and has been indeed calculated in \cite{BVE_PRB}) with a global response of the S/F system to the external magnetic field. Indeed the calculations of the total current in \cite{BVE_PRB} were made in a special gauge (vanishing vector-potential at S/F interface).  This gauge does not permit to determine the total response, because the presence of the superconductor should modify it. In other words, the full supercurrent induced inside the F film by the internal magnetic induction should be completely compensated by the screening currents inside the S layer. The applied external magnetic field is zero in such problem. As a result, we get the electromagnetic proximity effect predicted in our recent publication \cite{Mironov_APL}. The result obtained in \cite{BVE_PRB} relates only to the change of sign of the total current integrated over the ferromagnet and, thus, despite of its importance for further studies does not allow making unambiguous conclusions about the global response of S/F system to the external magnetic field.

In the above citation from our paper \cite{Mironov_2018} we are speaking of course about the global response of S/F system to the external magnetic field and, therefore, our citation of the paper \cite{BVE_PRB} in this context without the additional explanations is misleading.

\end{document}